\documentclass[titlepage,twoside,12pt]{article}
\usepackage{amssymb}
\usepackage{amsfonts}
\begin{document}

{\Large \bf What is wrong with von Neumann's \\ \\ theorem on "no hidden variables"} \\ \\

Elem\'{e}r E Rosinger \\
Department of Mathematics \\
and Applied Mathematics \\
University of Pretoria \\
Pretoria \\
0002 South Africa \\
eerosinger@hotmail.com \\ \\

{\bf Abstract} \\

It is shown that, although correct mathematically, the celebrated 1932 theorem of von Neumann
which is often interpreted as proving the impossibility of the existence of "hidden variables"
in Quantum Mechanics, is in fact based on an assumption which is physically not reasonable.
Apart from that, the alleged conclusion of von Neumann proving the impossibility of the
existence of "hidden variables" was already set aside in 1952 by the counterexample of the
possibility of a physical theory, such as given by what is usually called the "Bohmian
Mechanics". \\
Similar arguments apply to other two well known mathematical theorems, namely, of Gleason, and
of Kochen and Specker, which have often been seen as equally proving the impossibility of the
existence of "hidden variables" in Quantum Mechanics. \\

{\bf 1. Hidden variables describing the states of a quantum system} \\

The presentation in the sequel follows arguments in Hemmick [1,2], as well as in Manin
[pp. 82-95]. The main aim is to highlight the {\it simplicity} and {\it clarity} both in the
mathematical argument, and in the fact that, physically, one of the assumptions in von
Neumann's theorem is not reasonable. \\

In general, a physical theory is specifying {\it two basic entities}, namely, a {\it state
space} and a {\it dynamics} in it. \\

Since of concern here are finite and non-relativistic quantum systems, the state space will be
a Hilbert space ${\cal H} = {\cal L}^2 ( \mathbb{R}^{3 N} )$, where $N \geq 1$ is the number
of quantum particles in the system considered. Each state, therefore, is described by a so
called {\it wave function} $\psi \in {\cal H}$, which must be nonzero, and in addition, it can
be chosen arbitrary, modulo an nonzero constant $c \in \mathbb{C}$, that is, both $\psi$ and
$c \psi$ are supposed to describe the same quantum state. Consequently, we shall assume that
all such states $\psi$ are normalized, that is, we have $| |~ \psi ~| | = 1$. \\

As far as the dynamics is concerned, it is given either by the evolution of the wave function
according to the Schr\"{o}dinger equation when no measurement is performed, or according to
a possible so called "collapse of the wave function" which may occur upon the performance of a
measurement. However, regarding von Neumann's mentioned theorem, none of such dynamics is
directly involved. \\

Specific to Quantum Mechanics is the assumption that the states given by wave functions
$\psi \in {\cal H}$ {\it cannot} be observed directly. Consequently, the need for a {\it third
basic entity}, namely, the assumptions regarding the {\it ways} of observation and the {\it
results} of {\it measurement}. Here we shall only mention those assumptions which are relevant
in the context of von Neumann's theorem. Furthermore, for convenience, we shall make a number
of simplifying assumptions which, however, do not affect von Neumann's theorem. \\

The set ${\cal O}$ of {\it observables} is given by all Hermitian operators on the Hilbert
space ${\cal H}$ which is the state space. Every measurement corresponds to an observable $O
\in {\cal O}$, and the value of such a measurement can only be one of the eigenvalues $\mu \in
\mathbb{R}$ of $O$. If the quantum system is in the state $\psi \in {\cal H}$ when the
measurement is performed, then the probability that the particular eigenvalue $\mu$ will be
obtained as the value of the measurement is given by \\

(1.1)~~~~ $ |~ < \psi ~|~ \phi_\mu > ~|~^2 $ \\

where $\phi_\mu \in {\cal H}$ is supposed to be the eigenvector of $O$ which corresponds to
the eigenvalue $\mu$, thus we have $O |~ \phi_\mu > = \mu |~ \phi_\mu >$. \\
It follows that, if the quantum system is in the state $\psi \in {\cal H}$ when the measurement
corresponding to the observable $O$ is performed, then the {\it expectation} of the measured
value is \\

(1.2)~~~~ $ E_\psi ( O ) ~=~ < \psi ~|~ O ~|~ \psi > $ \\

The issue of "hidden variables" has arisen due to the fact that, in general, one cannot make a
more precise predictive statement about the outcome of a measurement than the probabilistic
one in (1.1), (1.2). Consequently, it can be suggested that the elements $\psi$ of the state
space ${\cal H}$ do not completely describe the state of the quantum system, and in order to
complete that description, one should add certain elements $\lambda$ which belong to a
suitable set $\Lambda$, and which describe the missing aspects of the quantum state. In other
words, the state space of the quantum system should be upgraded from ${\cal H}$ to the
Cartesian product ${\cal H} \times \Lambda$, and the states would now become pairs $( \psi,
\lambda )$, where $\psi \in {\cal H}$ are the usual wave functions, while $\lambda \in
\Lambda$ are the previously "hidden variables". \\
Von Neumann called "dispersion free" any such state $( \psi, \lambda )$. \\

Now the aim of the above is to obtain a "value function" \\

(1.3) ~~~~ $ V : {\cal H} \times \Lambda \times {\cal O} ~~\longrightarrow~~ \mathbb{R} $ \\

which eliminates the randomness or dispersion in (1.1), (1.2). It follows that for every state
$( \psi, \lambda )$ and observable $O$, we must have for the respective well determined value
the relation \\

(1.4) ~~~~ $ V ( \psi, \lambda, O ) ~=~ E_\psi ( O ) $ \\

and as noted by von Neumann, we must also have the property \\

(1.5) ~~~~ $ f ( V ( \psi, \lambda, O ) ) ~=~ V ( \psi, \lambda, f ( O ) ) $ \\

for every function $f : \mathbb{R} \longrightarrow \mathbb{R}$ which is defined by a
polynomial with real coefficients, thus (1.4) results in \\

(1.6) ~~~~ $ f ( E_\psi ( O ) ) ~=~ E _\psi ( f ( O ) ) $ \\ \\

{\bf 2. Von Neumann's Theorem} \\

The question asked by von Neumann was the following : what is the general form of a function \\

(2.1) ~~~~ $ E : {\cal H} \times {\cal O} ~~\longrightarrow~~ \mathbb{R} $ \\

which has the property, see (1.2) \\

(2.2) ~~~~ $ E ( \psi, O ) ~=~ E_\psi ( O ) ~=~ < \psi ~|~ O ~|~ \psi > $ \\

for every $\psi \in {\cal H},~ O \in {\cal O}$. \\

The assumptions on such a function $E$ made by von Neumann were the following {\it three} ones.
First \\

(2.3) ~~~~ $ E ( \psi, {\bf 1} ) ~=~ 1,~~~ \psi \in {\cal H} $ \\

where ${\bf 1} \in {\cal O}$ is the observable given by the identity operator on ${\cal H}$.
Thus all states $\psi \in {\cal H}$ are eigenvectors for ${\bf 1}$, and they all have the
eigenvalue $1 \in \mathbb{R}$. \\
Second, $E$ is real-linear in its second argument, namely \\

(2.4) ~~~~ $ E ( \psi, \alpha A + \beta B ) ~=~ \alpha E ( \psi, A ) + \beta E ( \psi, B ) $ \\

for $\psi \in {\cal H},~ A, B \in {\cal O},~ \alpha, \beta \in \mathbb{R}$. \\
Finally, for any observable $P \in {\cal O}$ which is a projection operator on ${\cal H}$, we
have \\

(2.5) ~~~~ $ E ( \psi, P ) ~\geq~ 0 $ \\

for all states $\psi \in {\cal H}$. \\

Under these assumption, we have \\ \\

{\bf Theorem 2.1 (von Neumann, 1932)} \\

The function $E$ must have the form \\

(2.6) ~~~~ $ E ( \psi, O ) ~=~ Tr ( U_\psi O ) $ \\

for $\psi \in {\cal H}$, where $U_\psi$ is a positive operator on ${\cal H}$, such that \\

(2.7) ~~~~ $ Tr ( U_\psi ) ~=~ 1 $ \\

{\bf Proof} \\

We extend the function $E$ in (2.1) to the set ${\cal L} ( {\cal H} )$ of all operators on
${\cal H}$, as follows. Given any operator $A \in {\cal L} ( {\cal H} )$, we can write in as \\

(2.8) ~~~~ $ A ~=~ A_{+} + i A_{-} $ \\

where $A_{+} = ( A + A^* ) / 2,~~ A_{-} = ( A - A^* ) / ( 2 i ) \in {\cal O}$ are Hermitian
operators, with $A^*$ denoting the Hermitian conjugate of $A$. \\
Now we can extend $E$ in (2.1) by defining \\

(2.9) ~~~~ $ E : {\cal H} \times {\cal L} ( {\cal H} ) ~~\longrightarrow~~ \mathbb{C} $ \\

where \\

(2.10) ~~~~ $ E ( \psi, A ) ~=~ E ( \psi, A_{+} ) + i E ( \psi, A_{-} ),~~~
                                                               \psi \in {\cal H} $ \\

It follows easily that this extended $E$ is complex-linear in its second argument. \\

In order to find out the form of $E$, we shall first consider operators $A \in {\cal L}
( {\cal H} )$ which have the particular form \\

(2.11) ~~~ $ A ~=~ \Sigma_{1 \leq n,m \leq M}~ |~ \phi_n >~< \phi_n ~|~ A ~|~
                                     \phi_m >~< \phi_m | $ \\

where $\phi_1,~.~.~.~ , \phi_N \in {\cal H}$ are orthonormal. Then the complex-linearity of
$E$ in (2.9) gives for $\psi \in {\cal H}$ the relation \\

(2.12) ~~~~ $ E ( \psi, A ) ~=~ \Sigma_{1 \leq n,m \leq M}~< \phi_n ~|~ A ~|~ \phi_m >
                           E ( \psi, | \phi_n > < \phi_m | ) $ \\

Now given a state $\psi \in {\cal H}$, we take the operator $U_\psi \in {\cal L} ( {\cal H} )$
as \\

(2.13) ~~~~ $ U_\psi ~=~ \Sigma_{1 \leq n,m \leq M}~ |~ \phi_n >~E ( \psi,
                                                 |~ \phi_n >~< \phi_m |~ )~ < \phi_m | $ \\

and obtain \\

(2.14) ~~~~ $ E ( \psi, A ) ~=~ Tr ( U_\psi A) $ \\

for all  $\psi \in {\cal H}$ and $A \in {\cal L} ( {\cal H} )$. \\

Let us show that $U_\psi$ is indeed a positive operator, for every state $\psi \in {\cal H}$.
Let $\chi \in {\cal H}$ be an arbitrary given state and $P_\chi \in {\cal L} ( {\cal H} )$ the
projection operator onto $\chi$. According to assumption (2.5), together with relation (2.14),
we have \\

(2.15) ~~~~ $ 0 ~\leq~ E ( \psi, P_\chi ) ~=~ Tr ( U_\psi P_\chi ) $ \\

However, this expression does not depend on the choice of the orthonormal states $\phi_1,
~.~.~.~ , \phi_N \in {\cal H}$. Therefore, we can assume that $\chi = \phi_1$, for instance.
In this case (2.13), (2.15) give \\

(2.16) ~~~~ $ 0 ~\leq~ E ( \psi, P_\chi ) ~=~ Tr ( U_\psi P_\chi ) ~=~
                                                    < \chi ~|~ U_\psi ~|~ \chi > $ \\

and since $\chi \in {\cal H}$ is arbitrary, it follows indeed that $U_\psi$ is a positive
operator. \\

At last, we show that (2.7) also holds. Indeed, from assumption (2.3), together with relation
(2.14), we have \\

(2.17) ~~~~ $ 1 ~=~ E ( \psi, {\bf 1} ) ~=~ Tr ( U_\psi {\bf 1} ) ~=~ Tr ( U_\psi ) $ \\

The general result follows now from a density argument, applied to (2.11). \\ \\

{\bf 3. Von Neumann's Argument Against "Hidden Variables"} \\

Let us consider observables given by one dimensional projection operators $P_\chi \in {\cal O}$
corresponding to states $\chi \in {\cal H}$. Then obviously \\

(3.1) ~~~~ $ P_\chi~^2 ~=~ P_\chi P_\chi ~=~ P_\chi $ \\

hence according to (1.5) and (2.2), it follows that \\

(3.2) ~~~~ $ ( E ( \psi, P_\chi ) )^2 ~=~ E ( \psi, P_\chi~^2 ) ~=~ E ( \psi, P_\chi ) $ \\

which means that \\

(3.3) ~~~~ $ \mbox{either}~~ E ( \psi, P_\chi ) ~=~ 0,~~~~\mbox{or}~~
                                                 E ( \psi, P_\chi ) ~=~ 1 $ \\

However, $E ( \psi, P_\chi )$ is obviously a continuous function in $\chi \in {\cal H}$. Thus
$E ( \psi, P_\chi )$ must be a constant, namely, with the value either $0$, or $1$. \\

If $E ( \psi, P_\chi ) = 0$, then (2.16) implies \\

(3.4) ~~~~ $ < \chi ~|~ U_\psi ~|~ \chi > ~=~ 0 $ \\

for $\chi \in {\cal H}$, with $||~ \chi ~|| = 1$, which means that $U_\psi = 0$, thus (2.7)
is contradicted. \\
If on the other hand $E ( \psi, P_\chi ) = 1$, then (2.16) yields \\

(3.5) ~~~~ $ < \chi ~|~ U_\psi ~|~ \chi > ~=~ 1 $ \\

for $\chi \in {\cal H}$, with $||~ \chi ~|| = 1$, which in view of (2.11), gives
$Tr ( U_\psi ) = M$, thus again contradicting (2.7). \\ \\

{\bf 4. Von Neumann's Physically Unrealistic Assumption} \\

As it turns out, in von Neumann's above theorem the assumption (2.4) on the real-linearity of
the function $E$ in (2.1) is physically questionable. Indeed, as it stands, this assumption is
required {\it regrdless} of whether the observables $A$ and $B$ are commuting, or not. \\

This issue of commutativity, however, is of crucial crucial importance. For if the observables
$A$ and $B$ are not commuting, then they cannot be measured simultaneously, thus the meaning
of the right hand term of (2.1) becomes unclear. \\

As an example, let us consider a spin $1/2$ quantum particle and measure it in three different
directions in the $xy$ plane, namely, along $x$, along $y$, and along the bisectrix $b$
between $x$ and $y$. Then the respective observables are related according to \\

(4.1) ~~~~ $ \sigma_b ~=~ ( \sigma_x + \sigma_y ) / \sqrt 2 $ \\

and their measurement can be performed by suitably oriented Stern-Gerlach magnets. However,
since $\sigma_x$ and $\sigma_y$ do not commute, thus the respective measurements are different
and distinct procedures which cannot happen simultaneously. It follows that the application of
condition (2.1) to (4.1) is questionable. And in fact, if one would do so, one would obtain
the {\it absurd} relation \\

(4.2) ~~~~ $ \pm 1 / 2 ~=~ ( \pm 1 / 2 \pm 1 / 2 ) / \sqrt 2 $ \\

since the eigenvalues of each of these observables are $\pm 1 / 2$. \\ \\

{~} \\ \\

{\large \bf Additional Comments} \\ \\

{\bf A : Two basic failures of existing mathematical models of} \\
\hspace*{0.7cm} {\bf Quantum Mechanics of nonrelativistic finite systems} \\

The first mathematical model for such quantum systems was set up in the late 1920s by J von
Neumann, see citation above. In this model one starts with the {\it state space} given by a
Hilbert space ${\cal L}^2(\mathbb{R}^d)$, with $1 \leq d < \infty$ related to the number of
quantum particles. The {\it observables} are typically densely defined unbounded closed
selfadjoint operators on the state space, and their real valued eigenvalues are supposed to be
the only quantities that can result from measurements. Till today, this first von Neumann
model is the one used in most of the first courses on Quantum Mechanics, with the frequent
simplification that the operators giving the observables are assumed to be in addition
everywhere defined and bounded. \\

Two of the basic observables in this model are the position $X$ and the momentum $P$. For ease
of notation we consider them when $d = 1$, thus the state space is simply
${\cal L}^2(\mathbb{R})$, and they then have the following form $X \psi (x) = x \psi (x)$ and
$P \psi (x) = (h/2 \pi i ) d/dx \psi (x)$, for $\psi \in {\cal L}^2(\mathbb{R})$ and $x \in
\mathbb{R}$. \\

Now, any better first course in Linear Functional Analysis will point out that {\it neither}
the position observable $X$, {\it nor} the momentum observable $P$ have eigenvectors in the
state space ${\cal L}^2(\mathbb{R})$. Thus they {\it cannot} have eigenvalues either, see for
instance the popular book of E Kreyszig, "Introductory Functional Analysis with Applications",
Wiley, New York, 1978, pp. 565 and 569. \\

This certainly {\it contradicts} the usual axioms of Quantum Mechanics, since it follows that
within the given quantum state space ${\cal L}^2(\mathbb{R})$, such basic observables like
position $X$ and momentum $P$ simply {\it cannot} be observed, as they {\it fail} to have
eigenvalues ! \\

So much for the extent to which the first von Neumann model satisfies the axioms of Quantum
Mechanics in their usual formulation, see for instance D T Gillespie, "A Quantum Mechanical
Primer", International Textbook Company, London, 1973. \\
Yet the fact remains that this first von Neumann model is most often used by physicists, and
also taught in most first courses in Quantum Mechanics. \\
Nevertheless, this failure is hardly ever mentioned. Thus one keeps endlessly facing this
basic failure, without being made explicitly and clearly aware of it. \\

Over the years, there have been some attempts to overcome this basic failure. The so called
"rigged Hilbert spaces", among others, where introduced for that purpose. However, such
attempts while solving some of the failures, happened to introduce other ones. \\

Of course, by using the Dirac {\it bra-ket} formalism, and with the help of the Dirac delta
function, certain purely symbolic or {\it formal manipulations} can be performed, in order to
try to paper over this basic failure. However, the fact remains that until now, that is, more
than seven decades after its first formulation, this failure of the first von Neumann model
has not been dealt with rigorously within the framework of the respective state spaces
${\cal L}^2(\mathbb{R}^d)$. \\

With respect to this omission we can mention as an example the recent book of S J Gustafson
and I M Sigal, "Mathematical Concepts of Quantum Mechanics", Springer Universitext, 2003. This
book, which dedicates its last chapter 17 to "Comments on Missing Topics ..." does not mention
anywhere the above basic failure. \\

Furthermore, one of the important reasons why the first von Neumann model is still so popular
is that, so far, it is the only one which allows for a simple, direct, explicit and
computational expression for the {\it space localization}. More precisely, this is a
localization in the {\it configuration space}. Namely, given any state, that is, any wave
function $\psi$ which is the solution of the Schr\"{o}dinger equation, or results following
the collapse caused by a measurement, then its space localization is done according to the Max
Born interpretation which says that $| \psi (x) |^2$ is the probability density of the event
that the quantum particle is in the neighbourhood of the configuration space point $x \in
\mathbb{R}$, when for instance $d = 1$. \\

Von Neumann himself was fully aware of the mentioned failure of his first model, and
consequently, suggested not much later a second model for Quantum Mechanics. In its present
day formulation, in this second von Neumann model one does {\it no longer} start with states
given by wave functions $\psi$, and instead, one starts with {\it observables} given by
elements $A$ in a suitable $C^*$-algebra, see K Hannabus, "An Introduction to Quantum Theory",
Oxford, 1997. \\

Unfortunately, this second von Neumann model suffers from another basic failure. And this also
holds for its various later, or even latest developments, such as for instance that suggested
in the 1960s by R Haag and D Kastler, or by R Haag in 1996. Namely, {\it none} of these type
of models can so far come anywhere near to a satisfactory expression of the {\it space
localization} of states which correspond to wave functions. And in this regard it is important
to remember that the space localization of wave functions, as given by the Max Born
interpretation, is a fundamental and unique feature of Quantum Mechanics, with no precedent
whatsoever in Classical Mechanics, including in General Relativity. \\

This second basic failure, this time of the second von Neumann model, remains for evermore
unmentioned in the literature, just like the first basic failure of the first von Neumann
model. An example regarding this omission is the recent book of R Haag, "Local Quantum Physics,
Fields, Particles, Algebras", Springer, 1996. \\

Once again, von Neumann was fully aware of such and other deficiencies of his first and second
models. Consequently, in less than a decade after suggesting his first model, he came up in
1936, together with George David Birkhoff, with the third model, namely, that of the so called
Quantum Logic. \\
This third model, however, was also found to be deficient. And after suggesting it in 1936,
and till his death in 1957, von Neumann was never to return again to the foundations of
Quantum Mechanics. \\

The {\it practical importance} of the above is as follows. Quantum Mechanics happens to be the
only basic theory of physics which does not have a rigorous enough mathematical model,
although it has several such models. \\
Physicists nevertheless manage to deal with this situation due to various purely formal or
symbolic devices and their manipulations, based on what usually goes by the name of "good
physical intuition". \\
On the other hand, as with the books of Gustafson and Sigal, or Haag, mathematicians end up by
being denied the knowledge that the existing models of Quantum Mechanics happen to have basic
failures. This situation, going on by now for nearly eight decades, is unfortunate. Indeed, in
modern times it has often happened that major mathematical theories were introduced and
developed being inspired by problems or difficulties in physics. On the other hand, and
needless to say, mathematical models of Quantum Mechanics which - as in the case of models of
all other basic theories of physics - would no longer exhibit basic failures like those two
mentioned above, may as well be to the benefit of physicists and of physics. \\ \\

{\bf B : The confession of John von Neumann} \\

As mentioned, between the late 1920s and 1936, von Neumann suggested no less than three
mathematical models for non-relativistic finite quantum systems. \\

The first one starts with the states given by a Hilbert space, and the observables given by
self-adjoint operators on that space. Generally, these operators are only densely defined and
unbounded, however, they are closed. \\
This model has an important advantage in that it allows for the Max Born interpretation
of the states, as given by the respective wave functions, in terms of probability densities on
the configuration space in which the quantum system is situated. This interpretation, called
usually {\it space localization}, is a {\it unique} and {\it distinctive} feature of Quantum
Mechanics, and it is particularly useful in a variety of considerations of theoretical or
practical nature. \\

The second model starts with observables, given by elements in a $C^*$-algebra, and then the
states are defined depending on the given $C^*$-algebra. This model does, so far, {\it not}
allow for a Max Born type interpretation of states. \\

The third model, suggested in 1936 in collaboration with George David Birkhoff, is only
concerned with the logical structure of the observables. In this model one does even less have
a Max Born type interpretation. \\

After that, till his death in 1957, von Neumann never returned to the issue of mathematical
modelling of quantum systems, at least not in his publications. \\

What is hardly known nowadays is that, in a letter to George David Birkhoff, von Neumann
wrote : \\

{\bf "I WOULD LIKE TO MAKE A CONFESSION WHICH MAY SEEM IMMORAL : I DO NOT BELIEVE
IN HILBERT SPACE ANYMORE."} \\

as quoted in : \\

G.D. Birkhoff, Proceedings of Symposia in Pure Mathematics, Vol. 2, p. 158, (Ed. R.P.
Dilworth), American Mathematical Society, Rhode Island, 1961. \\

And according to George David Birkhoff, the respective letter of von Neumann was dated 13
November, 1935.

\end{document}